\documentclass[runningheads]{llncs}
\usepackage[T1]{fontenc}
\usepackage[american]{babel}
\usepackage[nodayofweek]{datetime} 
\usepackage[dvipsnames]{xcolor}
\usepackage{graphicx}
\usepackage{tikz,pgfplots}
\usetikzlibrary{shapes.geometric, arrows, positioning, calc, backgrounds, trees, shadings}
\usepgfplotslibrary{colorbrewer}
\pgfplotsset{log ticks with fixed point, compat=1.16}
\usepackage{booktabs}
\usepackage{amsmath}  
\usepackage{amsfonts} %
\usepackage{url}
\usepackage{fancyvrb}   
\usepackage[nolist]{acronym}
\usepackage{hyperref}
\usepackage{placeins}
\usepackage[style=base,aboveskip=.3em,belowskip=.3em]{caption}
\usepackage{subcaption}
\usepackage[inline]{enumitem} 
\usepackage{comment}
\usepackage{xspace}
\usepackage[super]{nth} 
\usepackage{array} 
\usepackage{arydshln} 

\newcommand{\Regel}{\textsc{Regel}\xspace}
\newcommand{\Forest}{\textsc{Forest}\xspace}

\newcommand{\AlphaRegex}{\textsc{AlphaRegex}\xspace}

\newcommand{\lcnameref}[1]{%
\bgroup
\let\nmu\MakeLowercase
\nameref{#1}\egroup}
\newcommand{\nmu}{}

\definecolor{synthesizer-color}{HTML}{eeeeee}
\definecolor{enumerator-color}{HTML}{aaaaaa}

\tikzset{%
    square/.style={
    enumerator-color!30!black,
    rectangle,
    rounded corners,
    minimum width=1.9cm,
    minimum height=.8cm,
    text centered,
    draw=enumerator-color,
    fill=enumerator-color!20
  },
  nosquare/.style={
    synthesizer-color!40!black,
    rectangle,
    rounded corners,
    text centered
  },
  arrow/.style={
    synthesizer-color!30!black,
    thick,
    rounded corners=10,
    font=\footnotesize,
    ->,>=stealth, draw=gray
  },
  enclosing_square/.style={
    synthesizer-color!30!black,
    draw=synthesizer-color,
    very thick,
    dotted,
    fill=synthesizer-color!20
  },
  synth-node/.style={
    synthesizer-color!30!gray
  }
}

\definecolor{scatter-color}{HTML}{0026d1}
\definecolor{timeout-color}{HTML}{ba0039}
\definecolor{error-color}{HTML}{a30b00}
\newcommand{\compareTimesPlot}[4]{
\begin{tikzpicture}
    \begin{axis}[xmode=log, ymode=log, grid=major,
                xlabel = {\footnotesize #2},
                ylabel = {\footnotesize #4},
                xmax = 5000 ,ymax = 5000,
                width=.99\linewidth, height=.99\linewidth,
                clip marker paths=true]
    \addplot[only marks, mark=x, fill opacity=0.6, draw opacity=0.9, color=scatter-color] table[x=#1, y=#3, col sep=comma] {data/time_comparison_15.csv};
    \addplot[domain=0.1:3600,smooth,thin]{x};
    \addplot [dotted, thick, timeout-color, opacity=0.8] coordinates {(0.1,3600) (3600,3600)}
        node at (3,2500) {\tiny{3600 second timeout}};
    \addplot [dotted, thick, timeout-color, opacity=0.8] coordinates {(3600,0.1) (3600,3600)}
        node [rotate=-90] at (2500,3) {\tiny{3600 second timeout}};
    \end{axis}
\end{tikzpicture}
} 

\begin{document}




\title{\texorpdfstring{\Forest: An Interactive Multi-tree Synthesizer\\for Regular Expressions}{FOREST: An Interactive Multi-tree Synthesizer for Regular Expressions}}

\titlerunning{\Forest: An Interactive Multi-Tree Synthesizer for Regular Expressions}
\author{Margarida Ferreira\inst{1,2} \and
Miguel Terra-Neves\inst{2} \and
Miguel Ventura\inst{2} \and
In\^es Lynce\inst{1} \and
Ruben Martins\inst{3}}

\authorrunning{Ferreira et al.}

\institute{
INESC-ID, Instituto Superior Ténico, Universidade de Lisboa, Portugal\\
\email{\{margaridaacferreira,	ines.lynce\}@tecnico.ulisboa.pt}
\and
OutSystems, Portugal\\
\email{\{miguel.neves,	miguel.ventura\}@outsystems.com}
\and
Carnegie Mellon University, USA\\
\email{rubenm@cs.cmu.edu}}
\maketitle

\def\subsectionautorefname{section}
\def\subsectionname{section}
\def\subsubsectionautorefname{section}
\def\subsubsectionname{section}

\begin{acronym}
    \acro{DSL}{Domain Specific Language}
    \acro{CFG}{Context-Free Grammar}
    \acro{PBE}{Programming by Example}
    \acro{SyGuS}{Syntax-Guided Synthesis}
    \acro{CEGIS}{Counterexample Guided Inductive Synthesis}
    \acro{OGIS}{Oracle Guided Inductive Synthesis}
    \acro{SMT}{Satisfiability Modulo Theories}
    \acro{MaxSMT}{Maximum Satisfiability Modulo Theories}
\end{acronym}

\SaveVerb{date}|[0-9][0-9]/[0-9][0-9]/[0-9][0-9][0-9][0-9]|
\SaveVerb{date2}|[0-9]{2}/[0-9]{2}/[0-9]{4}|
\SaveVerb{date_day_mo_caps}|([0-9]{2})/([0-9]{2})/[0-9]{4}|
\SaveVerb{date_year_cap}|[0-9]{2}/[0-9]{2}/([0-9]{4})|

\begin{abstract}

Form validators based on regular expressions are often used on digital forms to prevent users from inserting data in the wrong format.
However, writing these validators can pose a challenge to some users.

We present \Forest, a regular expression synthesizer for digital form validations.
\Forest produces a regular expression that matches the desired pattern for the input values 
and 
a set of conditions over capturing groups that ensure the validity of integer values in the input.
Our synthesis procedure is based on enumerative search and uses a \acf{SMT} solver to explore and prune the search space.
We propose a novel representation for regular expressions synthesis, multi-tree, which induces patterns in the examples and uses them to split the problem through a divide-and-conquer approach.
We also present a new \ac{SMT} encoding to synthesize capture conditions for a given regular expression.
To increase confidence in the synthesized regular expression, we implement user interaction based on distinguishing inputs. 

We evaluated \Forest{} on real-world form-validation instances using regular expressions. Experimental results show that \Forest{} successfully returns the desired regular expression in 72\% of the instances and outperforms \Regel, a state-of-the-art regular expression synthesizer.

\end{abstract}

\section{Introduction}
Regular expressions (also known as regexes) are powerful mechanisms for describing patterns in text with numerous applications. 
One notable use of regexes is to perform real-time validations on the input fields of digital forms.
Regexes help filter invalid values, such as typographical mistakes (`typos') and format inconsistencies. 
Aside from validating the format of form input strings, regular expressions can be coupled with capturing groups.
A capturing group is a sub-regex within a regex that is indicated with parenthesis and 
captures the text matched by the sub-regex inside them.
Capturing groups are used to extract information from text and, in the domain of form validation, they can be used to enforce conditions over values in the input string.
In this paper, we focus on the capture of integer values in input strings, and we use the notation \(\$i, i \in \{0, 1, ...\}\) to refer to the integer value of the text captured by the \((i+1)\)\textsuperscript{th} group.

Form validations often rely on complex regexes which require programming skills that not all users possess.
%
To help users write regexes, prior work has proposed to synthesize regular expressions from natural language~\cite{Regel20,DBLP:conf/naacl/KushmanB13,DBLP:conf/emnlp/LocascioNDKB16,DBLP:conf/emnlp/ZhongGYPXLLZ18} or from positive and negative examples~\cite{Regel20,DBLP:conf/popl/Gulwani11,AlphaRegex16,Fidex16}. Even though these techniques assist users in writing regexes for search and replace operations, they do not specifically target digital form validation and do not take advantage of the structured format of the data.

In this paper, we propose \Forest{}, a new program synthesizer for regular expressions that targets digital form validations.
\Forest takes as input a set of examples and returns a regex validation that validates them. 
%
\Forest accepts three types of examples:
\begin{enumerate*}[label=(\roman*)]
    \item \textbf{valid examples}: correct values for the input field, 
    \item \textbf{invalid examples}: incorrect values for the input field due to their \textit{format}, and
    \item \textbf{conditional invalid examples} (optional): incorrect values for the input field due not to their format but to their \textit{values}.
\end{enumerate*}
\Forest outputs a regex validation, consisting of two components:
\begin{enumerate*}[label=(\roman*)]
    \item a \textbf{regular expression} that matches all valid and none of the invalid examples and
    \item \textbf{capture conditions} that express integer conditions for values in the examples that are satisfied by all valid but none of the conditional invalid examples.
\end{enumerate*}

\subsubsection{\nmu Motivating \nmu Example}\label{sec:motivating-example}
Suppose a user is writing a form where one of the fields is a date that must respect the format DD/MM/YYYY. The user wants to~accept:
\vspace{-2.2em}
\begin{multicols}{3}
    \begin{itemize}[label={},topsep=0em]
    \item 19/08/1996
    \item 26/10/1998
    \item 22/09/2000
    \item 01/12/2001
    \item 29/09/2003
    \item 31/08/2015
    \end{itemize}
\end{multicols}
\vspace{-1em}
\noindent
But not:
\vspace{-1em}
\begin{multicols}{3}
\begin{itemize}[label={},topsep=0em]
\item 19/08/96
\item 26-10-1998
\item 22.09.2000
\item 1/12/2001
\item 29/9/2003
\item 2015/08/31
\end{itemize}
\end{multicols}
\vspace{-1em}
\noindent
A regular expression can be used to enforce this format.
Instead of writing it, the user may simply use the two sets of values as \textit{valid} and \textit{invalid} input examples to \Forest{}, who will output the regex \UseVerb{date2}.


Additionally, suppose the user wants to validate not only the format, but also the values in the date.
We consider as \textit{conditional invalid} the examples:
\vspace{-.9em}
\begin{multicols}{3}
    \begin{itemize}[label={}]
    \item 33/08/1996
    \item 26/00/1998
    \item 22/13/2000
    \item 00/12/2001
    \item 12/31/2003
    \item 52/03/2015
    \end{itemize}
\end{multicols}
\vspace{-.9em}\noindent
To ensure only valid values are inserted as the day and month, \Forest outputs a regex validation complete with the necessary conditions over capturing groups:
\UseVerb{date_day_mo_caps}, \(\$0 \le 31 \wedge \$0 \ge 1 \wedge \$1 \le 12 \wedge \$1 \ge 1\).

\medskip

As we can see in the \lcnameref{sec:motivating-example}, data inserted into digital forms is usually structured and shares a common pattern among the valid examples.
In this example, the data has the shape \texttt{dd/dd/dddd} where \texttt{d} corresponds to a digit. This contrasts with general regexes for search and replace operations that are often performed over unstructured text. \Forest takes advantage of this structure by automatically detecting these patterns and using a divide-and-conquer approach to split the expression into simpler sub-expressions, solving them independently, and then merging their information to obtain the final regular~expression.
Additionally, \Forest computes a set of capturing groups over the regular expression, which it then uses to synthesize integer conditions that further constrain the accepted values for that form field.

Input-output examples do not require specialized knowledge and are accessible to users.
There is, however, one downside to using examples as a specification: they are ambiguous. There can be solutions that, despite matching the examples, do not produce the desired behavior in situations not covered in them.
The ambiguity of input-output examples raises the necessity of selecting one among multiple candidate solutions.
%
To this end, we incorporate a user interaction model based on distinguishing inputs
for both the synthesis of the regular expressions and the synthesis of the capture~conditions.

In summary, this paper makes the following contributions:
\begin{itemize}[topsep=.1ex]
    \item We propose a multi-tree SMT representation for regular expressions that leverages the structure of the input 
    to apply a  divide-and-conquer approach.
    \item We propose a new method to synthesize capturing groups for a given regular expression and integer conditions over the resulting captures.
    \item We implemented a tool, \Forest{}, 
    that interacts with the user to disambiguate the provided specification. \Forest{} is evaluated on real-world instances and its performance is compared with a state-of-the-art synthesizer.
\end{itemize}
\section{Synthesis Algorithm Overview}

\begin{figure}[t]
    \centering
    \includegraphics[scale=.25]{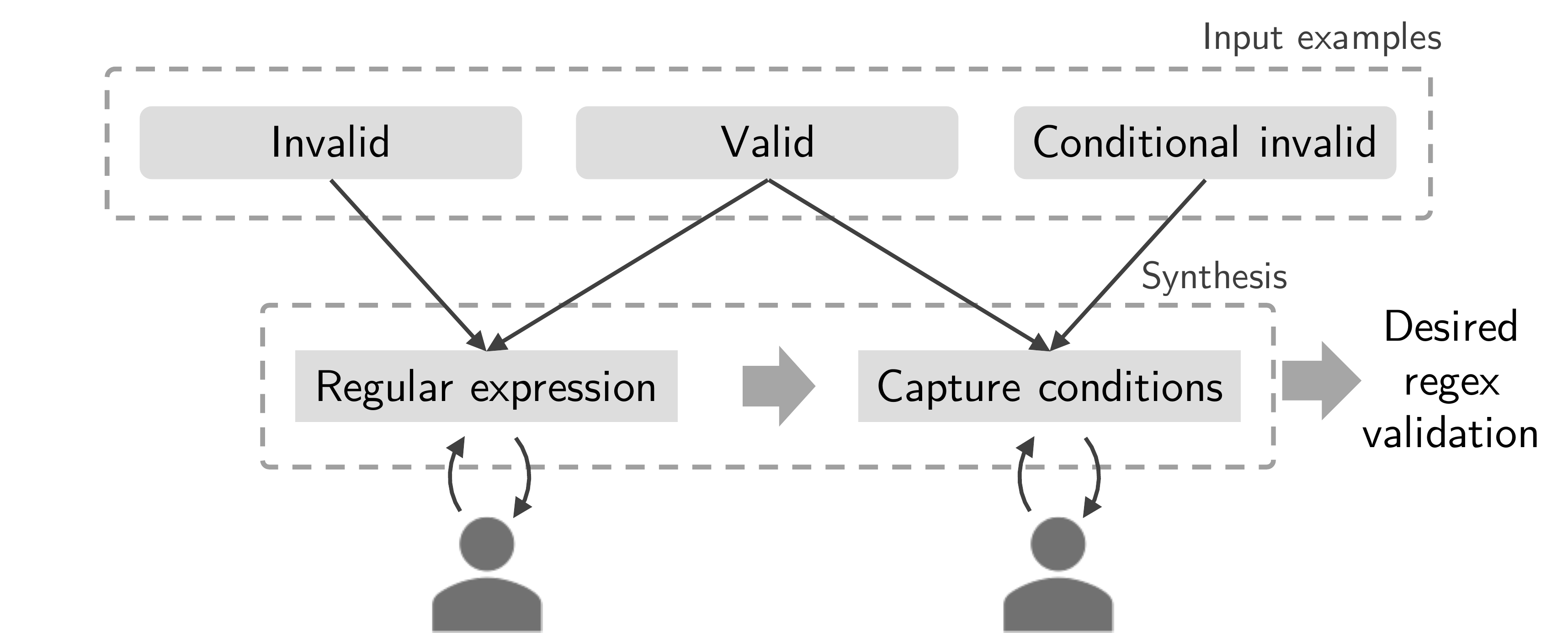}
    \captionsetup{belowskip=-.7em}
    \caption{Regex synthesis}
    \label{fig:regex-synthesis}
\end{figure}

\begin{figure}[t]
    \centering
    \includegraphics[scale=.25]{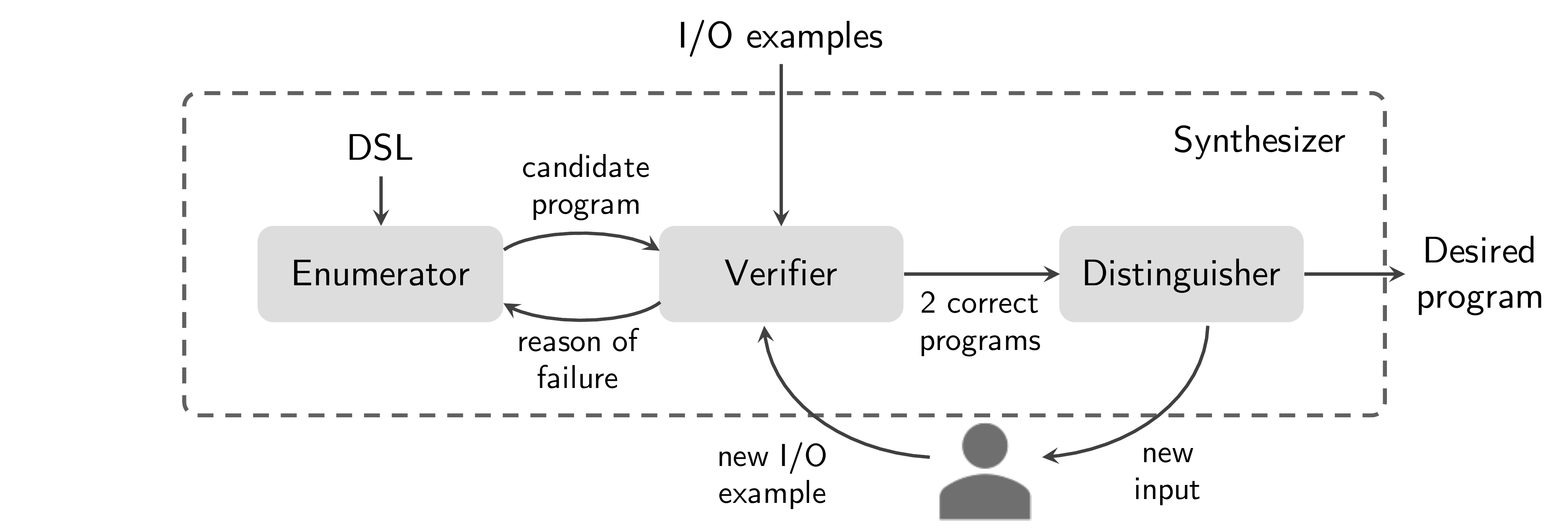}
    \captionsetup{belowskip=-.7em}
    \caption{Interactive enumerative search}
    \label{fig:interactive-synthesis}
\end{figure}

The task of automatically generating a program that satisfies some desired behavior expressed as a high-level specification is known as Program Synthesis. \acf{PBE} is a branch of Program Synthesis where the desired behavior is specified using input-output examples. 

Our synthesis procedure is split into two stages, each relative to an output component. 
First, \Forest synthesizes the regular expression, which is the basis for the synthesis of capturing groups.
Secondly, \Forest synthesizes the capture conditions, by first computing a set of capturing groups and then the conditions to be applied to the resulting captures.
The synthesis stages are detailed in sections \ref{sec:regex-synthesis} and \ref{sec:cap_groups_synthesis}.
\autoref{fig:regex-synthesis} shows the regex validation synthesis pipeline. 
Both stages of our synthesis algorithm employ enumerative search, a common approach to solve the problem of program synthesis \cite{DBLP:conf/pldi/FengMBD18,DBLP:conf/pldi/FengMGDC17,AlphaRegex16,Orvalho19,DBLP:conf/cav/ReynoldsBNBT19}.
The enumerative search cycle is depicted in~\autoref{fig:interactive-synthesis}.

There are two key components for program enumeration: the \textit{enumerator} and the \textit{verifier}.
The \textit{enumerator} successively enumerates programs from the a predefined \ac{DSL}. Following the Occam's razor principle, programs are enumerated in increasing order of complexity.
The \ac{DSL} defines the set of operators that can be used to build the desired program, as well as the values each operator can take as argument.
\Forest dynamically constructs its regular expression DSL based on the user-provided examples to fit the problem at hand: it is as restricted as possible, without losing the expressiveness necessary to ensure it includes the correct program. 
The regular expression DSL construction procedure is detailed in \autoref{sec:dsl}.
%

For each enumerated program, the \textit{verifier} subsequently checks whether it satisfies the provided examples.
Program synthesis applications generate very large search spaces; nevertheless, the search space can be significantly reduced by pruning several infeasible expressions along with each incorrect expression found. 
In the first stage of the regex validation synthesis, the enumerated programs are regular expressions.
The enumeration and pruning of regular expressions is described in \autoref{sec:enumeration}.
In the second stage of regex validation synthesis, we deal with the enumeration of capturing groups over a pre-existing regular expression. This process is described in \autoref{sec:cap_groups_enumeration}.

To circumvent the ambiguity of input-output examples, 
\Forest{} implements an interaction model.
A new component, the \textit{distinguisher}, ascertains, for any two given programs, whether they are equivalent.
When \Forest finds two different validations that satisfy all examples and are not equivalent, it creates a \textit{distinguishing input}: a new input that has a different output for each solution.
To disambiguate between two programs,
\Forest shows the new input to the user, who classifies it as valid or invalid, effectively choosing one program over the other.
The new input-output pair is added to the examples, and the enumeration process continues until there is only one solution left.
This interactive cycle is described for the synthesis of regular expressions in \autoref{sec:regex-interaction} and capture conditions in \autoref{sec:cap_groups_interaction}.

\section{Regular Expressions Synthesis}\label{sec:regex-synthesis}
In this section we describe the enumerative synthesis procedure that generates a regular expression that matches all valid examples and none of the invalid.

\subsection{Regular Expressions \texorpdfstring{\ac{DSL}}{DSL}}\label{sec:dsl}
Before the synthesis procedure starts, we define which operators can be used to build the desired regular expression and the values each operator can take as~argument. 
%
%
%
\Forest's regular expression \ac{DSL} includes the regex union and concatenation operators, as well as several regular expression quantifiers:
\begin{itemize}[topsep=0pt,leftmargin=*]
    \item \textit{Kleene} closure: \(r^*\) matches \(r\) zero or more times,
    \item positive closure: \(r^+\) matches \(r\) one or more times,
    \item option: \(r?\) matches \(r\) zero or one times,
    \item ranges: \(r\{m\}\) matches \(r\) exactly \(m\) times, and \(r\{m,n\}\) matches \(r\) at least \(m\) times and at most \(n\) times.
\end{itemize}

\noindent
The possible values for the range operators are limited depending on the valid examples provided by the user. For the single-valued range operator, \(r\{m\}\), we consider only the integer values such that \(2 \le m \le l\), where \(l\) is the length of the longest valid example string.
In the two-valued range operator, \(r\{m,n\}\), the values of \(m\) and \(n\) are limited to integers such that \(0 \le m < n \leq l\).
The tuple (0,1) is not considered, since it is equivalent to the option quantifier:~\(r\{0,1\} = r?\).

All operators can be applied to regex literals or composed with each other to form more complex expressions.
The regex literals considered in the synthesis procedure 
include the individual letters, digits or symbols present in the examples and all character classes that contain them.
The character classes contemplated in the DSL are \texttt{[0-9]}, \texttt{[A-Z]}, \texttt{[a-z]} and all combinations of those, such as \texttt{[A-Za-z]} or \texttt{[0-9A-Za-z]}. Additionally, \texttt{[0-9A-F]} and \texttt{[0-9a-f]} are used to represent hexadecimal numbers.


\begin{example}\label{ex:dsl}
%
Consider the valid examples in the \lcnameref{sec:motivating-example}.
%
The length of the longest valid example is \(l~=~10\), so the possible range values are \(m \in \{2, ..., 10\}\) for the single-argument range, and \((n, m): 0 \leq n < m \leq 10\) for the two-argument range. The characters in the valid examples are `/', which becomes itself a regex literal, and digits, which introduce the character class~\texttt{[0-9]}.
\end{example}


\subsection{Regex Enumeration}\label{sec:enumeration}

\begin{figure}[t]
\centering
\begin{tikzpicture}[
level distance=.7cm,
level 1/.style={sibling distance=5.0cm},
level 2/.style={sibling distance=3.0cm},
level 3/.style={sibling distance=1.3cm},
level 4/.style={sibling distance=.9cm,level distance=.5cm}]
\tikzstyle{every node}=[circle,draw, inner sep=.7mm]
\tikzstyle{every label}=[font=\small,rectangle, draw=none]

\node (Root) [label={concat}] {}
child {
    node [label={concat}] {} 
    child {
        node [label={concat}] {} 
        child {
            node [label={range}] {} 
            child { node [label=below:{\Verb![0-9]!}] {} }
            child { node [label=below:{2}] {} }
        }
        child {
            node [label={\texttt{/}}] {} 
            child { node [label={[text=gray]below:\(\epsilon\)}, draw=gray, right=.08cm] {} edge from parent[draw=gray] }
            child { node [label={[text=gray]below:\(\epsilon\)}, draw=gray, left=.08cm] {} edge from parent[draw=gray] }
        }
    }
    child {
        node [label={concat}] {} 
        child {
            node [label={range}] {} 
            child { node [label=below:{\Verb![0-9]!}] {} }
            child { node [label=below:{2}] {} }
        }
        child {
            node [label={\texttt{/}}] {} 
            child { node [label={[text=gray]below:\(\epsilon\)}, draw=gray, right=.08cm] {} edge from parent[draw=gray] }
            child { node [label={[text=gray]below:\(\epsilon\)}, draw=gray, left=.08cm] {} edge from parent[draw=gray] }
        }
    }
}
child {
    node [label={range}] {} 
    child {
        node [label={\Verb![0-9]!}, right=.3cm] {}
        child {
            node [label={[text=gray]:\(\epsilon\)}, draw=gray, right=0cm] {} edge from parent[draw=gray]
            child { node [label={[text=gray]below:\(\epsilon\)}, draw=gray, right=.08cm] {} edge from parent[draw=gray] }
            child { node [label={[text=gray]below:\(\epsilon\)}, draw=gray, left=.08cm] {} edge from parent[draw=gray] }
        }
        child {
            node [label={[text=gray]:\(\epsilon\)}, draw=gray, left=0cm] {} edge from parent[draw=gray]
            child { node [label={[text=gray]below:\(\epsilon\)}, draw=gray, right=.08cm] {} edge from parent[draw=gray] }
            child { node [label={[text=gray]below:\(\epsilon\)}, draw=gray, left=.08cm] {} edge from parent[draw=gray] }
        }
    }
    child {
        node [label={4}, left=.3cm] {}
        child {
            node [label={[text=gray]:\(\epsilon\)}, draw=gray, right=0cm] {} edge from parent[draw=gray]
            child { node [label={[text=gray]below:\(\epsilon\)}, draw=gray, right=.08cm] {} edge from parent[draw=gray] }
            child { node [label={[text=gray]below:\(\epsilon\)}, draw=gray, left=.08cm] {} edge from parent[draw=gray] }
        }
        child {
            node [label={[text=gray]:\(\epsilon\)}, draw=gray, left=0cm] {} edge from parent[draw=gray]
            child { node [label={[text=gray]below:\(\epsilon\)}, draw=gray, right=.08cm] {} edge from parent[draw=gray]}
            child { node [label={[text=gray]below:\(\epsilon\)}, draw=gray, left=.08cm] {} edge from parent[draw=gray]}
        }
    }
};

\end{tikzpicture}
\captionsetup{belowskip=-.7pt}
\caption{\UseVerb{date2} represented as a \(k\)-tree with \(k = 2\).}
\label{fig:date-ktree}
\end{figure}
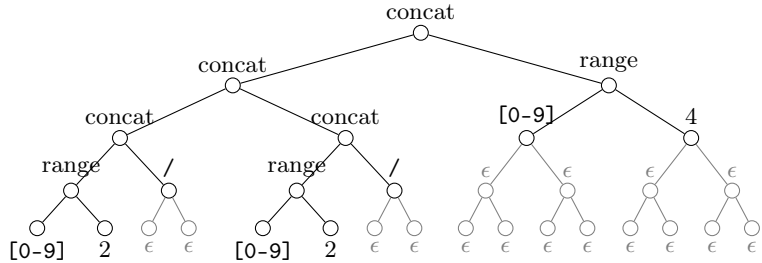{}

To enumerate regexes, the synthesizer requires a structure capable of representing every feasible expression.
We opt to use a tree-based representation of the search space, using \(k\)-trees.
A \(k\)-tree of depth \(d\) is a tree in which every internal node has exactly \(k\) children and every leaf node is at depth \(d\).
A program corresponds to an assignment of a DSL construct to each tree node.
The node's descendants are the construct's arguments.
If \(k\) is the greatest arity among all DSL constructs, then a \(k\)-tree of depth \(d\) can represent all programs of depth up to \(d\) in that \ac{DSL}.
The arity of constructs in \Forest{}'s regex \acp{DSL} is at most 2, so all regexes in the search space can be represented using 2\nobreakdash-trees.
To allow constructs with arity smaller than \(k\), some children nodes are assigned the \textit{empty} symbol, \(\epsilon\).
In \autoref{fig:date-ktree}, the regex from the \lcnameref{sec:motivating-example}, \UseVerb{date2}, is represented as a 2-tree of depth 5.

To explore the search space in order of increasing complexity, we enumerate \(k\)-trees of lower depths first and progressively increase the depth of the trees as previous depths are exhausted.
This way we ensure the first regex found is of the smallest depth possible.
The enumerator encodes the \(k\)-tree as an SMT formula,
whose constraints
%
%
ensure that the program is well-typed. 
A model that satisfies the formula represents 
a valid regex. Due to space constraints we omit the \(k\)-tree encoding but further details can be found
in the literature~\cite{DBLP:conf/sigsoft/ChenMF19,Orvalho19}.



\subsubsection{Multi-tree representation.}
We considered several validators for digital forms and observed that many regexes in this domain are the concatenation of relatively simple regexes. However, the successive concatenation of simple regexes quickly becomes complex in its \(k\)-tree representation.
Recall the regex for date validation presented in the \lcnameref{sec:motivating-example}: \UseVerb{date2}.
Even though this is the concatenation of 5 simple sub-expressions, each 
of depth at most~2, its representation as a {\(k\)-tree} has depth~5, as shown in \autoref{fig:date-ktree}.

The main idea behind the multi-tree constructs is to allow the number of concatenated sub-expressions to grow without it reflecting exponentially on the encoding. 
The multi-tree structure consists of \(n\) \(k\)-trees, whose roots are connected by an artificial root node, 
interpreted as an \(n\)-ary concatenation operator.
%
This way, we are able to represent regexes using fewer nodes. \autoref{fig:date-multitree} is the multi-tree representation of the same regex as \autoref{fig:date-ktree}, and shows that the multi-tree construct can represent this expression using half the~nodes.

\begin{figure}[t]
\centering
\definecolor{top-node-color}{HTML}{205090}
\begin{tikzpicture}[
level 1/.style={sibling distance=2cm, level distance=1cm},
level 2/.style={sibling distance=.7cm, level distance=.5cm},
empty/.style={edge from parent/.style={solid,gray,draw}}]
\tikzstyle{every node}=[circle,draw=black, solid, inner sep=.8mm, edge from parent/.style={solid,gray,draw=black}]
\tikzstyle{edge from parent}=[draw=black, solid]
\tikzstyle{every label}=[rectangle, draw=none]

\node (Root) [label={[text=top-node-color]:concat}, draw=top-node-color,thick, dotted] {} 
child {
    node [label={range}] {} edge from parent[draw=top-node-color,dashed]
    child {
        node [label=below:{\Verb![0-9]!}, draw=black] {} }
    child {
        node [label=below:{2}] {} }
}
child {
    node [label={\texttt{/}}] {} edge from parent[draw=top-node-color,dashed]
    child[empty] {
        node [label={[text=gray]below:\(\epsilon\)}, draw=gray] {}}
    child[empty] {
        node [label={[text=gray]below:\(\epsilon\)}, draw=gray] {}}
}
child {
    node [label={range}] {} edge from parent[draw=top-node-color,dashed]
    child {
        node [label=below:{\Verb![0-9]!}] {} }
    child {
        node [label=below:{2}] {} }
}
child {
    node [label={\texttt{/}}] {} edge from parent[draw=top-node-color,dashed]
    child {
        node [label={[text=gray]below:\(\epsilon\)}, draw=gray] {}  edge from parent[draw=gray] }
    child {
        node [label={[text=gray]below:\(\epsilon\)}, draw=gray] {}  edge from parent[draw=gray] }
}
child {
    node [label={range}] {}edge from parent[draw=top-node-color,dashed]
    child {
        node [label=below:{\Verb![0-9]!}] {} }
    child {
        node [label=below:{4}] {} }
};

\end{tikzpicture}
\captionsetup{belowskip=-.7em}
\caption{\UseVerb{date2} represented as a multi-tree with \(n=5\) and \(k=2\), resulting from the concatenation of 5 simpler regexes.}
\label{fig:date-multitree}
\end{figure}
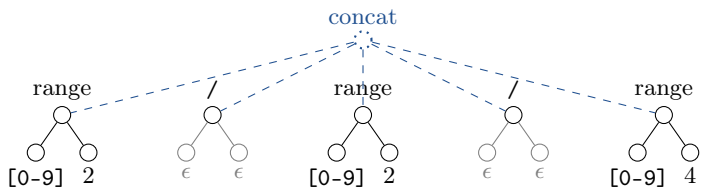{}

The \(k\)-tree enumerator
successively explores \(k\)-trees of increasing depth.
However, multi-tree has two measures of complexity: the depth of the trees, \(d\), and the number of trees, \(n\).
\Forest{} employs two different methods for increasing these values: static multi-tree and dynamic multi-tree.

\subsubsection{Static multi-tree.} \label{sec:static-multi-tree} In the static multi-tree method, the synthesizer fixes \(n\) and progressively increases \(d\). To find the value of \(n\), there is a preprocessing step, in which \Forest{} identifies patterns in the valid examples. 
This is done by first identifying substrings common to all examples.
A substring is considered a dividing substring if it occurs exactly the same number of times and in the same order in all examples.
Then, we split each example before and after the dividing substrings. Each example becomes an array of \(n\) strings.

\begin{example}\label{ex:splitting}
Consider the valid examples from the \lcnameref{sec:motivating-example}.
In these examples, `/' is a dividing substring because it occurs in every example, and exactly twice in each one. `0' is a common substring but not a dividing substring because it does not occur the same number or times in all examples. After splitting on `/', each example becomes a tuple of 5 strings:
\vspace{-.9em}
\begin{multicols}{2}
    \begin{itemize}[label={},nosep]
    \item (`19', `/', `08', `/', `1996')
    \item (`26', `/', `10', `/', `1998')
    \item (`22', `/', `09', `/', `2000')
    \item (`01', `/', `12', `/', `2001')
    \item (`29', `/', `09', `/', `2003')
    \item (`31', `/', `08', `/', `2015')
    \end{itemize}
\end{multicols}
\end{example}
\noindent
Then, we apply the multi-tree method with \(n\) trees. For every \(i \in \{1, ..., n\}\), the \(i^{th}\) sub-tree represents a regex that matches all strings in the \(i^{th}\) position of the split example tuples and the concatenation of the \(n\) regexes will match the original example strings.
Since each tree is only synthesizing a part of the original input strings, a reduced DSL is recomputed for each~tree. 

\subsubsection{Dynamic multi-tree.}\label{sec:dynamic-multi-tree} The dynamic multi-tree method is employed when the examples cannot be split because there are no dividing substrings.
In this scenario, the enumerator will still use a multi-tree construct to represent the regex. However, the number of trees is not fixed and all trees use the original, complete DSL. 
A multi-tree structure with \(n\) \(k\)-trees of depth \(d\) has \(n \times (k^d - 1)\) nodes.
\Forest enumerates trees with different values of \((n, d)\) in increasing order of number of nodes, starting with \(n = 1\) and \(d = 2\), 
a simple \(k\)-tree of depth 2.

\subsubsection{Pruning.}\label{sec:pruning}
%
There can be multiple regular expressions that, although syntactically different, have the same semantics.
%
%
\noindent
To prevent the enumeration of equivalent regular expressions,
we add SMT constraints that block all but one possible representation of each regex. Take, for example, the equivalence \({(r?)+ \equiv r*}\). We want to consider only one way to represent this regex, so we add a constraint to block the construction \((r?)+\) for any regex \(r\).
%
Another such equivalence results from the idempotence of union: \(r|r = r\).
To prevent the enumeration of expressions of the type \(r|r\), every time the union operator is assigned to a node~\(i\),
we force the sub-tree underneath \(i\)'s left child to be different from the sub-tree underneath \(i\)'s right child by at least one node.
%
When we enumerate a regex that is not consistent with the examples, it is eliminated from the search space. 
Along with the incorrect regex, we want to eliminate regexes that are equivalent to it. The union operator in the regular expressions DSL is commutative: \(r|s = s|r\), for any regexes \(r\) and \(s\).
Thus, whenever an expression containing \(r|s\) is discarded, we eliminate the expression that contains \(s|r\) in its place as well.
\subsection{Regex Disambiguation}\label{sec:regex-interaction}
To increase confidence in the synthesizer's solution, \Forest{} disambiguates the specification by interacting with the user. 
We employ an interaction model based on distinguishing inputs, which has been successfully used in several synthesizers  \cite{DBLP:journals/pvldb/LiCM15,DBLP:conf/sigmod/WangCB17,DBLP:conf/pldi/WangCB17,DBLP:conf/uist/MayerSGLMPSZG15}.
To produce a distinguishing input, we require an SMT solver with a regex theory, such as Z3 \cite{z3,z3str317}.
Upon finding two regexes that satisfy the user-provided examples, \(r_1\) and \(r_2\), we use the SMT solver to solve the~formula:
\begin{equation}\label{eq:distinguishing}
  \exists s: r_{1}(s) \neq r_{2}(s),
\end{equation}
where \(r_{1}(s)\) (resp. \(r_{2}(s)\)) is True if and only if \(r_{1}\) (resp. \(r_{2}\)) matches the string~\(s\). The string~\(s\) that satisfies \eqref{eq:distinguishing} is a distinguishing input. \Forest{} asks the user to classify this input as valid or invalid.
\(s\) is added to the set of valid or invalid examples accordingly, effectively eliminating either \(r_1\) or \(r_2\) from the search space.
After the first interaction with the user, the synthesis procedure continues only until the end of the current depth and number of trees.

\section{Capturing Groups Synthesis}\label{sec:cap_groups_synthesis}

In this section we describe the enumerative synthesis procedure that generates the second component of the regex validation:
a set of integer conditions over captured values that are satisfied by all valid examples but not by any of the conditional invalid examples.
%
    
\subsection{Capturing Groups Enumeration}\label{sec:cap_groups_enumeration}

To enumerate capturing groups, \Forest starts by identifying the regular expression's atomic sub-regexes:
the smallest sub-regexes whose concatenation results in the original complete regex.
For example, \verb![0-9]{2}! is an atomic sub-regex: there are no smaller sub-regexes whose concatenation results in it.
It does not make sense to place a capturing group inside atomic sub-regexes: \verb!([0-9]){2}! does not have a clear meaning.
Once identified, the atomic sub-regexes are placed in an ordered list. 
Enumerating capturing groups over the regular expression is done by enumerating non-empty disjoint sub-lists of this list. The elements inside each sub-list form a capturing~group.

\begin{example}
Recall the date regex: \UseVerb{date2}. The respective list of atomic sub-regexes is [\verb![0-9]{2}!, \verb!/!, \verb![0-9]{2}!, \verb!/!, \verb![0-9]{4}!].
%
The following are examples of sub-lists of the atomic sub-regexes list and their resulting capturing groups:

\centering
\setlength{\extrarowheight}{5pt}\small
\begin{tabular}{rcl}
[[\verb![0-9]{2}!], \verb!/!, \verb![0-9]{2}!, \verb!/!, \verb![0-9]{4}!] & $\rightarrow$ & \verb!([0-9]{2})/[0-9]{2}/[0-9]{4}! \\

[[\verb![0-9]{2}!], \verb!/!, [\verb![0-9]{2}!], \verb!/!, [\verb![0-9]{4}!]] & $\rightarrow$ & \verb!([0-9]{2})/([0-9]{2})/([0-9]{4})!
\end{tabular}
\end{example}

\subsection{Capture Conditions Synthesis}
To compute capture conditions, we need all conditional invalid examples to be matched by the regular expression.
%
After, \Forest enumerates capturing groups as described in \autoref{sec:cap_groups_enumeration}. The number of necessary capturing groups is not known beforehand, so we enumerate capturing groups in progressively increasing~number.
%
%
%
A capture condition is a 3-tuple: it contains the captured text, an integer comparison operator and an integer argument.
\Forest considers only two integer comparison operators, $\le$ and $\ge$. However, the algorithm can be easily expanded to include other operators.
%
Let \(\mathcal{C}\) be a set of capturing groups and \(\mathcal{C}(x)\) the integer captures that result from applying \(\mathcal{C}\) to example string \(x\).
Let \(\mathcal{D}_\mathcal{C}\) be the set of all possible capture conditions over capturing groups \(\mathcal{C}\).
\(\mathcal{D}_\mathcal{C}\) results from combining each capturing group with each integer operator.
%
%
Finally, let \(\mathcal{V}\) be the set of all valid examples, \(\mathcal{I}\) the set of all conditional invalid examples, and \(\mathcal{X} = \mathcal{V} \cup \mathcal{I}\) the union of these 2 sets.

Given capturing groups \(\mathcal{C}\), \Forest uses \ac{MaxSMT} to select from \(\mathcal{D}_\mathcal{C}\) the minimum set of conditions that are satisfied by all valid examples and none of the conditional invalid.
To encode the problem, we define two sets of Boolean variables. 
First, we define \(s_{\textit{cap}, x}\) for every \(\textit{cap} \in \mathcal{C}(x)\) and \(x \in \mathcal{X}\).
\(s_{\textit{cap}, x} =\) True if capture \textit{cap} in example \(x\) satisfies all used conditions that refer to it.
We also define \(u_{\textit{cond}}\) for all \(\textit{cond} \in \mathcal{D}_\mathcal{C}\).
\(u_{\textit{cond}} = \) True means condition \textit{cond} is used in the solution.
Additionally, we define a set of integer variables \(b_{\textit{cond}}\), for all conditions \(\textit{cond} \in \mathcal{D}_\mathcal{C}\) that represent the integer argument present in each condition.

Let \(\textrm{SMT}(\textit{cond}, x)\) be the \ac{SMT} representation of the condition \(\textit{cond}\) for example \(x\):
the capture is an integer value, 
and the integer argument is the corresponding \(b_{\textit{cond}}\) variable.
%
%
Let \(\mathcal{D}_{cap} \subseteq \mathcal{D}_\mathcal{C}\) be the set of capture conditions that refer to capture \textit{cap}. Constraint \eqref{eq:cap_cond_s_def} states that a capture \textit{cap} in example \(x\) satisfies all conditions if and only if for every condition that refers to \textit{cap}
either it is not used in the solution or it is satisfied for the value of that capture in that example:
\begin{equation}\label{eq:cap_cond_s_def}
    s_{cap,x} \leftrightarrow \bigwedge_{cond \in \mathcal{D}_{\textit{cap}}} u_{cond} \rightarrow \textrm{SMT}(\textit{cond}, x).
\end{equation}

\begin{example}
Recall the first valid string from the \lcnameref{sec:motivating-example}: \(x_0\) = ``27/10/2017''.
Suppose \Forest has already synthesized the desired regular expression and enumerated a capturing group that corresponds to the day:\\ \verb`([0-9]{2})/[0-9]{2}/[0-9]{4}`. 
Let \(\textit{cond}_0\) and \(\textit{cond}_1\) be the conditions that refer to the first (and only) capturing group, \(\$0\), and operators \(\le\) and \(\ge\) respectively. 
The SMT representation for \(\textit{cond}_0\) 
and \(x_0\) is
\(\textrm{SMT}(\textit{cond}_0, x_0)\; =\;  27 \le b_{\textit{cond}_0}\). 
%
Constraint \eqref{eq:cap_cond_s_def}~is:
\begin{equation*}
    s_{0,x_0} \leftrightarrow (u_{\textit{cond}_0} \rightarrow 27 \le b_{\textit{cond}_0}) 
    \land (u_{\textit{cond}_1} \rightarrow 27 \ge b_{\textit{cond}_1}).
\end{equation*}
\end{example}

%
%
Then, we ensure the used conditions are satisfied by all valid examples and none of the conditional invalid examples:

\begin{equation}\label{eq:cap_cond_a_clause}
    \bigwedge_{x \in \mathcal{V}} \;\bigwedge_{cap \in \mathcal{C}(x)} s_{cap,x} \;\land\; \bigwedge_{x \in \mathcal{I}} \;\bigvee_{cap \in \mathcal{C}(x)} \neg s_{cap,x}.
\end{equation}

Since we are looking for the minimum set of capture conditions, 
we add soft clauses to penalize the usage of capture conditions in the solution:
\begin{equation}
    \bigwedge_{cond \in \mathcal{D}_\mathcal{C}} \neg\, u_{cond}.
\end{equation}

We consider part of the solution only the capture conditions whose \(u_{cond}\) is True in the resulting \ac{SMT} model. We also extract the values of the integer arguments in each condition from the model values of the \(b_{\textit{cond}}\) variables.

\subsection{Capture Conditions Disambiguation}\label{sec:cap_groups_interaction}

To ensure the solution meets the user's intent, \Forest disambiguates the specification using, once again, a procedure based on distinguishing inputs.
%
%
%
Once \Forest finds two different sets of capture conditions \(\mathcal{S}_1\) and \(\mathcal{S}_2\)
that satisfy the specification, we look for a distinguishing input: a string \(c\) which satisfies all capture conditions in \(\mathcal{S}_1\), but not those in \(\mathcal{S}_2\), or vice-versa. First, to simplify the problem, \Forest eliminates from \(\mathcal{S}_1\) and \(\mathcal{S}_2\) conditions which are present in both: these are not relevant to compute a distinguishing input. Let \(\mathcal{S}_1^*\) (resp. \(\mathcal{S}_2^*\)) be the subset of \(\mathcal{S}_1\) (resp. \(\mathcal{S}_2\)) containing only
the distinguishing conditions, i.e.,
the conditions that differ from those in  \(\mathcal{S}_2\) (resp. \(\mathcal{S}_1\)).

We do not compute the distinguishing string \(c\) directly. Instead, we compute the integer value of the distinguishing captures in \(c\), i.e., the captures that result from applying the regular expression and its capturing groups to the distinguishing input string. 
We define \(|\mathcal{C}|\) integer variables, \(c_i\), which correspond to the values of the distinguishing captures:
\(c_0, c_1, ..., c_{|\mathcal{C}|} = \mathcal{C}(c).\)


As before, let \(\textrm{SMT}(\textit{cond}, c)\) be the \ac{SMT} representation of each condition \(\textit{cond}\). Each capture in \(\mathcal{C}(c)\) is represented by its respective \(c_i\), the operator maintains it usual semantics and the integer argument is its value in the solution to which the condition belongs.
Constraint \eqref{eq:cap_cond_distinguishing_constraint} states that \(c\) satisfies the conditions in one solution but not the~other.

\begin{equation}\label{eq:cap_cond_distinguishing_constraint}
    \bigwedge_{\textit{cond} \,\in\, \mathcal{S}_1^*} \textrm{SMT}(\textit{cond}, c) \quad\ne\; \bigwedge_{\textit{cond} \,\in\, \mathcal{S}_2^*} \textrm{SMT}(\textit{cond}, c).
\end{equation}

In the end, to produce the distinguishing string \(c\), \Forest picks an example from the valid set, applies the regular expression with the capturing groups to it, and replaces its captures with the model values for \(c_i\).

\Forest{} asks the user to classify \(c\) as valid or invalid.
Depending on the user's answer, \(c\) is added as a valid or conditional invalid example, effectively eliminating either \(S_1\) or \(S_2\) from the search space.

\begin{example}\label{ex:cond_cap_keep_distinct}
Recall the examples from the \lcnameref{sec:motivating-example}. No example invalidates a date with the day 32, so \Forest will find two correct sets of capture conditions over the regular expression \verb`([0-9]{2})/([0-9]{2})/[0-9]{4}`: \(\mathcal{S}_1 = \{\$0 \le 31, \$0 \ge 1, \$1 \le 12, \$1 \ge 1\}\), and \(\mathcal{S}_2 = \{\$0 \le 32, \$0 \ge 1, \$1 \le 12, \$1 \ge 1\}\).
First, we define to sets containing only the distinguishing captures: \(\mathcal{S}_1^* = \{\$0 \le 31\}\) and \(\mathcal{S}_2^* = \{\$0 \le 32\}\).
Then, to find \(c_0\), the value of the distinguishing capture for these solutions, we solve the constraint:
\[\exists c_0 : c_0 \le 31 \ne c_0 \le 32\]
and get the value \(c_0 = 32\) which satisfies \(\mathcal{S}_2^*\) (and \(\mathcal{S}_2\)), but not \(\mathcal{S}_1^*\) (or \(\mathcal{S}_1\)).

If we pick the first valid example, ``19/08/1996'' as basis for \(c\), the respective distinguishing input is \(c\)~=~``32/08/1996''. Once the user classifies \(c\) as invalid, \(c\) is added as a conditional invalid example and \(\mathcal{S}_2\) is removed from consideration.
\end{example}

\section{Related Work}
\label{sec:related}

Program synthesis has been successfully used in many domains such as string processing~\cite{DBLP:conf/ijcai/KiniG15,DBLP:conf/aaai/RazaG17,DBLP:conf/popl/Gulwani11,Fidex16}, query synthesis~\cite{DBLP:journals/pvldb/LiCM15,DBLP:conf/pldi/WangCB17,Orvalho19}, data wrangling~\cite{DBLP:conf/sigsoft/ChenMF19,DBLP:conf/pldi/FengMGDC17}, and functional synthesis~\cite{DBLP:conf/cp/FedyukovichG19,golia2020manthan}.
In this section, we discuss prior work on the synthesis of regular expressions~\cite{AlphaRegex16,Regel20} that is most closely related to our approach.

Previous approaches that perform general string processing~\cite{DBLP:conf/popl/Gulwani11,Fidex16} restrict the form of the regular expressions that can be synthesized. In contrast, we support a wide range of regular expressions operators, including the Kleene closure, positive closure, option, and range. 
More recent work that targets the synthesis of regexes is done by \AlphaRegex~\cite{AlphaRegex16} and \Regel~\cite{Regel20}. \AlphaRegex performs an enumerative search and uses under- and over-approximations of regexes to prune the search space. However, \AlphaRegex is limited to the binary alphabet and does not support the kind of regexes that we need to synthesize for form validations.
\Regel~\cite{Regel20} is a state-of-the-art synthesizer of regular expressions based on a multi-modal approach that combines input-output examples with a natural language description of user intent. They use natural language to build hierarchical sketches that capture the high-level structure of the regex to be synthesized.
In addition, they prune the search space by using under- and over-approximations and symbolic regexes combined with SMT-based reasoning. \Regel's evaluation~\cite{Regel20} has shown that their PBE engine is an order of magnitude faster than \AlphaRegex. While \Regel targets more general regexes that are suitable for search and replace operations, we target regexes for form validation which usually have more structure.
In our approach, we take advantage of this structure to split the problem into independent subproblems. This can be seen as a special case of sketching~\cite{DBLP:journals/sttt/Solar-Lezama13} where each hole is independent. Our pruning techniques are orthogonal to the ones used by \Regel and are based on removing equivalent regexes prior to the search and to remove equivalent failed regexes during search.
To the best of our knowledge, no previous work focused on the synthesis of conditions over capturing groups.

Instead of using input-output examples, there are other approaches that synthesize regexes solely from natural language~\cite{DBLP:conf/naacl/KushmanB13,DBLP:conf/emnlp/LocascioNDKB16,DBLP:conf/emnlp/ZhongGYPXLLZ18}. We see these approaches as orthogonal to ours and expect that \Forest can be improved by hints provided by a natural language component such as was done in \Regel.
\section{Experimental Results}

\paragraph{Implementation.}
\Forest{} is implemented in Python 3.8 on top of \textsc{Trinity}, a general-purpose synthesis framework \cite{trinity19}.
All \ac{SMT} formulas are solved using the Z3 SMT solver, version  4.8.9 \cite{z3}. 
To find distinguishing inputs in regular expression synthesis, \Forest uses Z3's theory of regular expressions (part of the theory of strings \cite{z3str317}).
To check the enumerated regexes against the examples, we use Python's regex library. \cite{PythonRe}
The results presented herein were obtained using an Intel(R) Xeon(R) Silver 4110 CPU @ 2.10GHz, with 64GB of RAM, running Debian GNU/Linux~10. All processes were run with a time limit of one hour.

\paragraph{Benchmarks.} To evaluate \Forest{}, we used 64 benchmarks based on real-world form-validation regular expressions. These were collected from regular expression validators in validation frameworks and from \texttt{regexlib}~\cite{regexlib}, a website where users can upload their own regexes. Among these 64 benchmarks there are different formats: national IDs, identifiers of products, date and time, vehicle registration numbers, postal codes, email and phone numbers.
For each benchmark, we generated a set of string examples.
All 64 benchmarks require a regular expression to validate the examples, but only 7 require capture conditions.
On average, each instance is composed of 13.2 valid examples (ranging from 4 to 33) and 9.3 invalid (ranging from 2 to 38). The 7 instances that target capture conditions have on average 6.3 conditional invalid examples (ranging from 4 to 8). 

\medskip\noindent
The goal of this experimental evaluation is to answer the following questions:
\begin{enumerate}[topsep=.1em, label=\textbf{Q\arabic*}.]
\item How does \Forest{} compare against \Regel? (\autoref{sec:comp-regel})
\item How does pruning affect multi-tree's time performance? (\autoref{sec:pruning-split})
\item How does 
static multi-tree improve on dynamic multi-tree? (\autoref{sec:pruning-split})
\item How does multi-tree compare against other enumeration-based encodings? (\autoref{sec:multi-tree-vs-encodings})
%
%
\item How many examples are required to return an accurate solution? (\autoref{sec:fewer-exs})
\end{enumerate}

\begin{table}[t]
\setlength{\dashlinedash}{.5ex}
\setlength{\dashlinegap}{1ex}
\setlength{\tabcolsep}{2ex}
\centering
\caption{Comparison of time performance using different synthesis methods.}
\begin{tabular}{@{}lcccc@{}}
\toprule
\textbf{Timeout (s)}  & \textbf{10} & \textbf{60} & \textbf{3600} \\ \midrule
\Forest{} (with interaction)   & 31 & 39 & 48   \\ 
\Forest's \nth{1} regex (no interaction)  & 40 & 46 & 50   \\ \hdashline
Multi-tree w/o pruning   & 20 & 32 & 38   \\
Dynamic-only multi-tree  & 5  & 10 & 18   \\ \hdashline
\(k\)-tree               & 4  & 9  & 15   \\
Line-based (w/o pruning) & 4  & 4  & 12   \\ \hdashline
\Regel{}                 & 28 & 38 & 47   \\
\Regel{} PBE             & 3  & 5  & 23   \\ \bottomrule
\end{tabular}
\label{table:number-solved}
\end{table}
\begin{figure}[t]
  \centering
  \begin{tikzpicture}
  \begin{axis}[
    xmode=linear,  ymode=linear,
    width=.53\linewidth, height=.47\linewidth,
    grid=major,
    ymax=3600 ,xmax=64, xmin=1, ymin=0.05,
    xtick distance=8, ytick distance=600,
    ylabel = {\footnotesize{Time (s)}},
    xlabel = {\footnotesize{}{Instances solved}},
    tick label style = {font=\scriptsize},
    legend style={
      anchor=west,
      at={(1.05,.5)}
    },
    cycle list/Dark2-8,
    cycle multi list={Dark2-8}  
  ]
  
    \addplot+[line width=.7pt, mark=square*, mark size=1.2] table [x=ranking, y=time, col sep=comma] {data/ranking_lines.csv};
    \addlegendentry{\footnotesize Line-based};
    
    \addplot+[line width=.7pt, mark=+] table [x=ranking, y=time, col sep=comma] {data/ranking_ktree.csv};
    \addlegendentry{\footnotesize \(k\)-tree};
    
    \addplot+[line width=.7pt, mark=diamond*] table [x=ranking, y=time, col sep=comma] {data/ranking_dynamic.csv};
    \addlegendentry{\footnotesize Dynamic multi-tree};
    
    \addplot+[line width=.7pt, mark=x] table [x=ranking, y=time, col sep=comma] {data/ranking_regel_no_sketches.csv};
    \addlegendentry{\footnotesize \Regel PBE};
    
    \addplot+[line width=.7pt, mark=otimes*, mark size=1.2] table [x=ranking, y=time, col sep=comma] {data/ranking_nopruning.csv};
    \addlegendentry{\footnotesize Multi-tree w/o pruning};
    
    \addplot+[line width=.7pt, mark=asterisk] table [x=ranking, y=time, col sep=comma] {data/ranking_regel_skcomp.csv};
    \addlegendentry{\footnotesize \Regel};
  
    \addplot+[line width=.7pt, mark=triangle*, mark size=1.5] table [x=ranking, y=time, col sep=comma] {data/ranking_forest.csv};
    \addlegendentry{\footnotesize \Forest};

    \addplot+[line width=.7pt, mark=Mercedes star] table [x=ranking, y=time, col sep=comma] {data/ranking_forest_first_regex.csv};
    \addlegendentry{\footnotesize \Forest's \nth{1} regex};
  
\end{axis}
\end{tikzpicture}
\captionsetup{belowskip=-.7em}
\caption{Instances solved using different methods}
\label{fig:comparison_all_methods}
\end{figure}
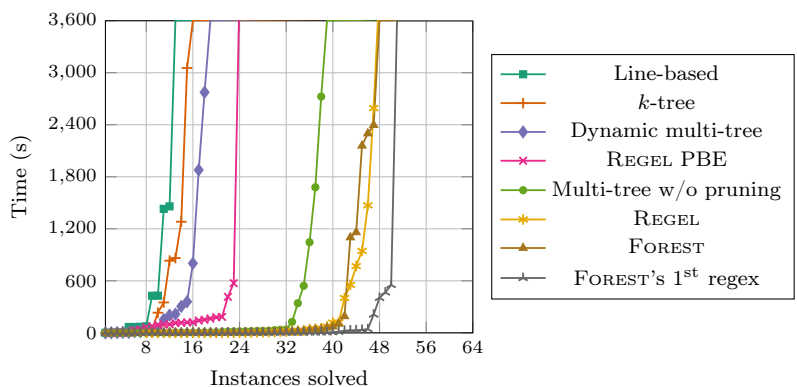

\Forest{}, by default, uses static multi-tree (when possible) with pruning. It correctly solves 31 benchmarks (48\%) in under 10 seconds.
In one hour, \Forest{} solves 48 benchmarks (75\%), with 96\% accuracy: only two solutions did not correspond to the desired regex validation. \Forest disambiguates only among programs at the same depth, and so if the first solution is not at the same depth as the correct one, the correct solution is never found.
After 1~hour of running time, \Forest{} is interrupted, but it prints its current best validation before terminating.
After the timeout, \Forest{} returned 2 more regexes, both the correct solution for the benchmark. \Forest{} never uses more than 500MB of RAM in each benchmark.
In all benchmarks to which \Forest returns a solution, the first matching regular expression is found in under 10 minutes. In 40 benchmarks, the first regex is found in under 10 seconds.  The rest of the time is spent disambiguating the input examples. 
\Forest interacts with the user to disambiguate the examples in 27 benchmarks. Overall, it asks 1.8 questions and spends 38.6 seconds computing distinguishing inputs, on average. 

Regarding the synthesis of capture conditions, in 5 of the benchmarks, we need only 2 capturing groups and at most 4 conditions. In these instances, the conditions' synthesis takes under 2 seconds. The remaining 2 benchmarks need 4 capturing groups and take longer: 99 seconds to synthesize 4 conditions and 1068 seconds for 6 conditions. During capture conditions synthesis, \Forest interacts 7.14 times and takes 0.1 seconds to compute distinguishing inputs, on~average.

\autoref{table:number-solved} shows the number of instances solved in under 10, 60 and 3600 seconds using \Forest{}, as well as using the different variations of the synthesizer which will be described in the following sections. The cactus plot in \autoref{fig:comparison_all_methods} shows the cumulative synthesis time on the y-axis plotted against the number of benchmarks solved by each variation of \Forest{} (on the x-axis). The synthesis methods that correspond to lines more to the right of the plot are able to solve more benchmarks in less time. 
We also compare solving times with \Regel~\cite{Regel20}.
\Regel takes as input examples and a natural description of user intent. We consider not only the complete \Regel synthesizer, but also the PBE engine of \Regel by itself, which we denote by \Regel PBE.

\subsection{Comparison with \Regel}\label{sec:comp-regel}
As mentioned in \autoref{sec:related}, \Regel's synthesis procedure is split into two steps: sketch generation (using a natural language description of desired behavior) and sketch completion (using input-output examples).
To compare \Regel and \Forest{}, we extended our 64 form validation benchmarks with a natural language description.
To assess the importance of the natural language description, we also ran \Regel using only its PBE engine. 
Sketch generation took on average 60 seconds per instance, and successfully generated a sketch for 63 instances. The remaining instance was run without a sketch.
We considered only the highest ranked sketch for each instance.
In \autoref{table:number-solved} we show how many instances can be solved with different time limits for sketch completion; note that these values do not include the sketch generation time.
\Regel{} returned a regular expression for 47 instances and timed out after one hour in the remaining 17.
Since \Regel does not implement a disambiguation procedure, the returned regular expression does not always exhibit the desired behavior, even though it correctly classifies all examples.
Of the 47 synthesized expressions, 31 exhibit the desired~intent. 
This is a 66\% accuracy, which is much lower than \Forest's at 96\%.
We also observe that \Regel's performance is severely impaired when using only its PBE~engine. 

51 out of the 63 generated sketches are of the form \(\square\{S_1, ..., S_n\}\),
where each \(S_i\) is a concrete sub-regex, i.e., has no holes.
This construct indicates the desired regex must contain \textit{at least} one of \(S_1, ..., S_n\),
and contains no information about the top-level operators that are used to connect them.
22 of the 47 synthesized regexes are based on sketches of that form, and they result from the direct concatenation of \textit{all} components in the sketch. No new components are generated during sketch completion.
Thus, most of \Regel{}'s sketches could be integrated into \Forest{}, whose multi-tree structure holds precisely those top-level operators that were missing from \Regel's~sketches.


\subsection{Impact of pruning the search space and splitting examples}\label{sec:pruning-split}
\begin{figure}[t]
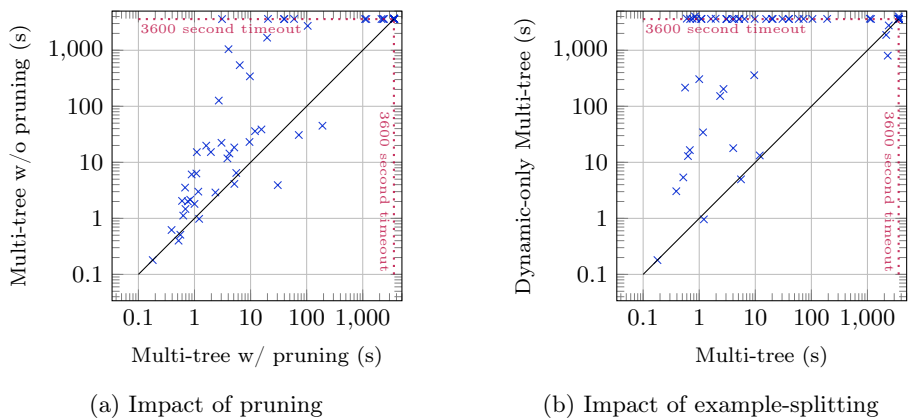

    \centering
    
    \begin{subfigure}[t]{0.45\textwidth}
         \centering
         \compareTimesPlot{multitree}{Multi-tree w/ pruning (s)}{nopruning}{Multi-tree w/o pruning (s)}
         \caption{Impact of pruning}
         \label{fig:mt_vs_nopruning}
     \end{subfigure}
     \hfill
     \begin{subfigure}[t]{0.45
\textwidth}
         \centering
         \compareTimesPlot{multitree}{Multi-tree (s)}{dynamic}{Dynamic-only Multi-tree (s)}
         \caption{Impact of example-splitting}
         \label{fig:mt_vs_dynamic}
     \end{subfigure}
\captionsetup{belowskip=-.7em}
\caption{Comparison of synthesis time using different variations of \Forest{}.}
\label{fig:time-comparison-funny}
\end{figure}

To evaluate the impact of pruning the search space as described in \autoref{sec:pruning}, we ran \Forest{} with all pruning techniques disabled.
In the scatter plot in \autoref{fig:mt_vs_nopruning}, we can compare the solving time on each benchmark with and without pruning. Each mark in the plot represents an instance. The value on the y-axis shows the synthesis time of multi-tree with pruning disabled and the value on the x-axis the synthesis time with pruning enabled. The marks above the \(y = x\) line (also represented in the plot) represent problems that took longer to synthesize without pruning than with pruning.
On average, with pruning, \Forest{} can synthesize regexes in 42\% of the time and enumerates about 15\% of the regexes before returning. There is no significant change in the number of interactions before returning the desired solution.

\Forest{} is able to split the examples and use static multi-tree as described in \autoref{sec:static-multi-tree} in 52 benchmarks (81\%). The remaining 12 are solved using dynamic multi-tree. To assess the impact of using static multi-tree we ran \Forest{} with a version of the multi-tree enumerator that does not attempt to split the examples, and jumps directly to dynamic multi-tree solving.
In the scatter plot in \autoref{fig:mt_vs_dynamic}, we compare the solving times of each benchmark.
Using static multi-tree when possible, \Forest{} requires, on average, less than two thirds of the time (59.1\%) to return the desired regex for the benchmarks solved by both methods.
Furthermore, static multi-tree allows \Forest{} to synthesize more complex expressions: the maximum number of nodes in a solution returned by dynamic multi-tree is 12 (average 6.7), while complete multi-tree synthesizes expressions of up to 24 nodes (average 10.3).

\subsection{\texorpdfstring{Multi-tree versus \(k\)-tree and line-based encodings}{Multi-tree versus k-tree and line-based encodings}}\label{sec:multi-tree-vs-encodings}

To evaluate the performance of the multi-tree encoding, we ran \Forest{} with two other enumeration encodings: \(k\)-tree and line-based.
The latter is a state of the art encoding for the synthesis of SQL queries \cite{Orvalho19}. \(k\)-tree is implemented as the default enumerator in \textsc{Trinity}~\cite{trinity19}, while the line-based enumerator is used as available in \textsc{Squares}~\cite{squares-webpage}. The \(k\)-tree encoding has a very similar structure to that of multi-tree, so our pruning techniques were easily applied to this encoding. On the other hand, line-based encoding is intrinsically different, making the pruning techniques harder to implement. Because of this, we compare line-based encoding to multi-tree without pruning.
In every other aspect, the three encodings were run in the same conditions, using \Forest{}'s regex \ac{DSL}.
\(k\)-tree is able to synthesize programs with up to 10 nodes, while the line-based encoding synthesizes programs of up to 9 nodes. Neither encoding~outperforms~multi-tree.

As seen in \autoref{table:number-solved}, line-based encoding does not outperform the tree-based encodings for the domain of regular expressions while it was much better for the domain of SQL queries~\cite{Orvalho19}. We conjecture the reason behind this disparity arises from the different nature of \ac{DSL}s. Most SQL queries, when represented as a tree, leave many branches of the tree unused and have more children per node, which results in a much larger tree and SMT encoding.


\subsection{Impact of fewer examples} \label{sec:fewer-exs}

To assess the impact of providing fewer examples on the accuracy of the solution, we ran \Forest{} with modified versions of each benchmark.
First, each benchmark was run with at most 10 valid and 10 invalid examples, chosen randomly among all examples. Conditional invalid examples are already very few per instance, so these were not altered. The accuracy of the returned regexes is slightly~lower.

With only 10 valid and 10 invalid examples, \Forest{} returns the correct regex in 93.5\% of the benchmarks, which represents a decrease of only 2.5\% relative to the results with all examples.
We also saw an increase in the number of interactions before returning, since fewer examples are likely to be more ambiguous. With only 10 examples, \Forest{} interacts on average 2.2 times per benchmark, which represents an increase of about a fifth.
The increase in the number of interactions reflects on a small increase in the synthesis time (less than 1\%).

After, we reduced the number of examples even further: only 5 valid and 5 invalid. The accuracy of \Forest{} in this setting was reduced to 71\%. On average, it interacted 4.3 times per benchmark, which is over two times more than~before. 


\section{Conclusions and Future Work}
Regexes are commonly used to enforce patterns and validate the input fields of digital forms. However, writing regex validations requires specialized knowledge that not all users possess.
We have presented a new algorithm for synthesis of regex validations from examples that leverages the common structure shared between valid examples. Our experimental evaluation shows that the multi-tree representation synthesizes three times more regexes than previous representations in the same amount of time and, together with the user interaction model, \Forest{} solves 72\% of the benchmarks with the correct user intent. We verified that \Forest{} maintains a very high accuracy with
as few as 10 examples of each kind.
We also observed that our approach outperforms \Regel{}, a state-of-the-art synthesizer, in the domain of form~validations.

As future work, we would like to explore the synthesis of more complex capture conditions, such as conditions depending on more than one capture. This would allow more restrictive validations; for example, in a date, the possible values for the day could depend on the month.
Another possible extension to \Forest is to automatically separate invalid from conditional invalid examples, making this distinction imperceptible to the user.
\bibliographystyle{splncs04}
\bibliography{references}
\end{document}